# A pseudo–binary interdiffusion study in the β–Ni(Pt)Al phase

Perumalsamy Kiruthika and Aloke Paul

Dept. of Materials Engineering, Indian Institute of Science, Bangalore, India

*Corresponding author: E-mail: aloke@materials.iisc.ernet.in, Tel.: 918022933242, Fax: 91802360 0472

**Abstract**

Interdiffusion study is conducted in the Ni-rich part of the β–Ni(Pt)Al phase following the pseudo-binary approach. Interdiffusion coefficients over the whole composition range considered in this study increases with the increase in Pt content, which is in line with the theoretical study predicting the decrease in vacancy formation and migration energy because of Pt addition. The trend of change in diffusion coefficient with the increase in Ni and Pt content indicates that Pt preferably replaces Ni antisites.

**Keywords:** diffusion, intermetallics, defects, pseudo–binary approach

## 1. Introduction

Bond coat is an integral part in jet engine applications for the protection of the base material *i.e.* Ni-based superalloys from oxidation. β–Ni(Pt)Al is one of the bond coats used preferably on the Ni-based superalloys. During service, an $Al_2O_3$ layer grows on top of it thereby protecting the base metal from oxidation. Bond coat acts as a reservoir for the continuous supply of Al such that a protective layer of alumina can grow immediately after spallation because of the thermal stress at the bond coat/alumina interface. Addition of Pt in β–NiAl increased the service life of turbine blades by manifold. Although the mechanism by which Pt provides beneficial effect is not well understood, it is found that Pt addition decreases the segregation of S at the bond coat and $Al_2O_3$ interface. In addition, it is believed that the diffusion rate of components increases to facilitate higher growth rate of the oxide layer by supplying Al at higher rate. This finds support in the theoretical analysis by Marino and Carter [1]. However, to the best of our knowledge, no relevant experimental studies are available in the literature. Minamino et al. [2] estimated the diffusion rate of Pt in different β-NiAl alloys. However, it should be noted here that the diffusion rates of Ni and Al because of presence of Pt is more important to study.

Therefore, the aim of this present study is to conduct diffusion couple experiments examining the role of Pt on the diffusion rates of Ni and Al based on quantitative diffusion analysis. A pseudo-binary approach seems to be a suitable technique for this purpose [3]. Added advantage of this method is that it mimics the composition profile of the bond coat in real application which develops by the interdiffusion between the bond coat and the superalloy [4–7].

## 2. Experimental procedure

Ni (99.9 wt.%), Al (99.9 wt.%) and Pt (99.99 wt.%) were used to prepare the alloys for making diffusion couple. For the purpose of making pseudo−binary diffusion couples two sets of alloys, $Ni_{60-x}Pt_xAl_{40}$ and $Ni_{50-x}Pt_xAl_{50}$, x = 5, 10 and 15 (all in atomic percentage), were melted in an arc melting furnace under Ar atmosphere. Average deviation of the compositions from intended ones was around ±1 at.%. These were homogenized in a vacuum furnace (~$10^{-4}$ Pa) at 1200 ºC for 100 h and the average compositions were measured randomly at different places of the blocks in an electron probe micro−analyzer (EPMA). Diffusion couples were prepared such that Pt remains constant in both side of the end members varying only Ni and Al. These experiments were conducted at 1100 ºC for 25 h. After the experiments, the samples were cross−sectioned and metallographicaly prepared for EPMA analysis.

## 3. Results and discussion

Till date, several experimental interdiffusion studies are conducted in the binary Ni−Al system [8−14]. However, as mentioned already the role of Pt on interdiffusion of Ni and Al are not examined. For the sake of comparison both binary and ternary (by pseudo-binary approach) experiments are conducted. In this, Ni−rich part of the β−Ni(Pt)Al is studied because of relevance to the application. Figure 1, shows the composition profiles developed in $Ni_{60}Al_{40}/Ni_{50}Al_{50}$ and $Ni_{50}Pt_{10}Al_{40}/Ni_{40}Pt_{10}Al_{50}$ diffusion couples during annealing. To apply pseudo−binary approach it is important that Pt has almost constant concentration over the whole interdiffusion zone, which is the case in these experiments. Reason for finding this behavior could be understood from Figure 2, which shows the variation of activity of components for $Ni_{50}Pt_{10}Al_{40}/Ni_{40}Pt_{10}Al_{50}$ across the interdiffusion zone extracted using CALPHAD. Activity for Ni and Al varies significantly, whereas, it is more or less constant for Pt.

In a binary system, the interdiffusion flux of a component at different compositions can be estimated by [3, 15, 16]

$$\tilde{J}_i = -\frac{\Delta N_i}{2t}\left[(1-Y_i)\int_{x^{-\infty}}^{x^*}\frac{Y_i}{V_m}dx + Y_i\int_{x^*}^{x^{+\infty}}\frac{(1-Y_i)}{V_m}dx\right] \quad (1)$$

where, $\tilde{J}_i$ is the interdiffusion flux. Minus sign comes from the fact that the element $i$ diffuses from right to left. $x^*$ is the position of interest. $Y_i = \frac{N_i - N_i^-}{N_i^+ - N_i^-}$ (where, $\Delta N_i = N_i^+ - N_i^-$) is the composition normalized variable introduced by Wagner [15], $N_i$ is the composition of the element $i$ in terms of atomic fraction or mol fraction. $-\infty$ and $+\infty$ are the left– and right– hand side of the unaffected parts of the diffusion couple. $V_m$ is the molar volume.

Following, the interdiffusion coefficients at the point of interest can be estimated by

$$\tilde{J}_i = -\tilde{D}\frac{dC_i}{dx} \quad (2)$$

where $C_i = N_i / V_m$ is the concentration. In a binary system, one can consider any of the components. The same estimation procedure can be followed in the pseudo– binary diffusion couple. In this case, as explained in Ref. [3], the interdiffusion coefficients can be calculated directly from the Al composition profile. On the other hand, (Ni+Pt) composition profile can also be used for the estimation of diffusion coefficients, since Pt shares the same sublattice as Ni [17]. Molar volume variation at different Pt content are estimated using the lattice parameter data available in literature [18], which were, in fact, determined from the alloys used in this work.

These are found to be similar to the data available in another reference [19]. The variations of the molar volumes for different Pt contents are shown in Figure 3.

The estimated interdiffusion coefficients are shown in Figure 4a. For comparison, two data sets estimating the interdiffusion coefficients in the binary Ni−Al system are incorporated [9, 13], in which the molar volume variation was used for the estimation. It can be seen that the data estimated in the present study are very close to them. Another interesting point to be noted here is that, as shown in Figure 4b, with increase in Pt content, there is significant increase in the interdiffusion coefficient. Interestingly, interdiffusion coefficient increases at higher rate as the composition moves towards the stoichiometric 50(Ni+Pt):50 Al composition.

This trend indicates the change in concentration of the defects assisting the diffusion process in a certain way. First of all, the interdiffusion coefficient increases at all compositions with the increase in Pt content. Theoretical analysis by Marino and Carter [1] indicates that the defect formation energy and migration energy decreases with the increase in Pt content leading to increase in diffusion coefficient. Secondly, it is a known fact that in the Ni–rich side of the β−Ni(Pt)Al phase, diffusion of both the components is assisted by the presence of Ni antisites [1]. The concentration of Ni antisites increases with the increase in deviation towards Ni–rich (Al–lean) side to compensate the deviation from the stoichiometric composition. Higher the deviation means higher the concentration of these defects leading to higher rate of diffusion of both the components, as it was found in interdiffusion studies [9, 14, 20]. The increase in Ni diffusion rate with increase in Ni content in the Ni–rich side of the β−NiAl was also found based on Ni tracer diffusion studies [21, 22]. Therefore, interdiffusion coefficient increases with the increase in Ni content. Additionally, as shown in Figure 4b with increase in Pt content the rate of increase of the interdiffusion coefficient is lesser on the Ni–rich (Al–lean) side than the ones near the stoichiometric composition. There could be two reasons behind this. It is well possible that the effect of Pt on increasing defect concentrations is higher in the Ni rich side. On the other hand, it is possible that Pt replaces some amount of Ni antisites since Pt shares the same sublattice [23]. Effect of alloying on defect concentration and diffusion rate was established before in another system. Rietveld analysis of the Nb–Si–Mo alloy conducted by Li et al. [24] shows that Mo has the site preference for the Si antisites in the $Nb_5Si_3$ phase. On the other hand, Si is the main diffusing species for the growth of this phase. Therefore, addition of Mo in this phase leads to the decrease in growth rate and the interdiffusion coefficient [25].

## 4. Conclusion

The role of Pt addition on diffusing components of Ni and Al in β–NiAl is examined. Estimated interdiffusion coefficients indicate that the diffusion rate increases with the increase in Pt content. Our results are in line with the theoretical predictions of Ref. [1] where it was obtained that the defect formation energy as well as the migration energy decreases with the increase in Pt content. The rate of increase in interdiffusion coefficient because of Pt addition decreases with the increase in Ni content (decrease in Al content), which indicates that Pt preferably replaces Ni antisites.

**Acknowledgement:** We would like to acknowledge the financial support from ARDB, India.

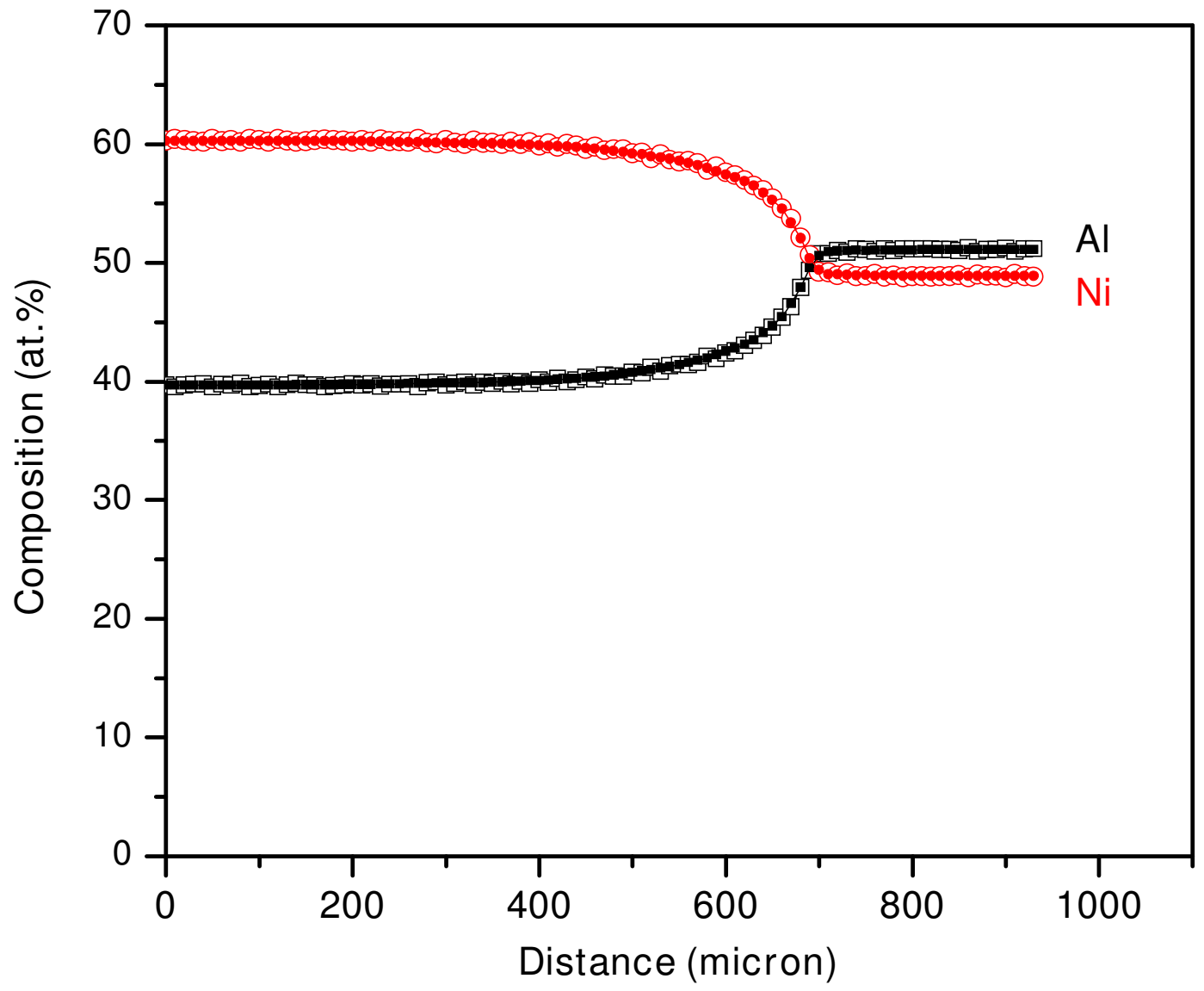

(a)

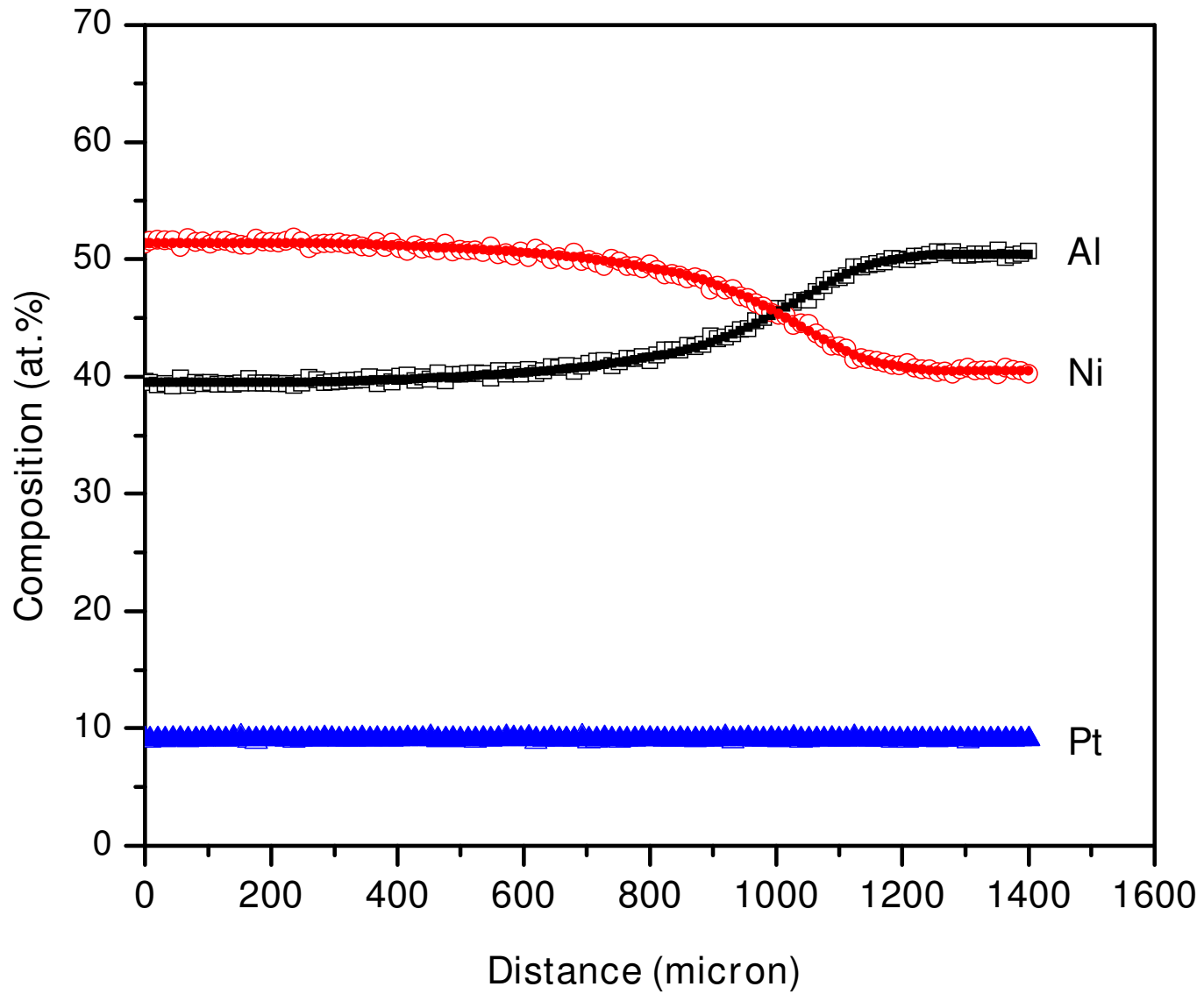

(b)

Fig 1: Compositional profiles developed in (a) $Ni_{60}Al_{40}/Ni_{50}Al_{50}$

(b) $Ni_{50}Pt_{10}Al_{40}/Ni_{40}Pt_{10}Al_{50}$ diffusion couples.

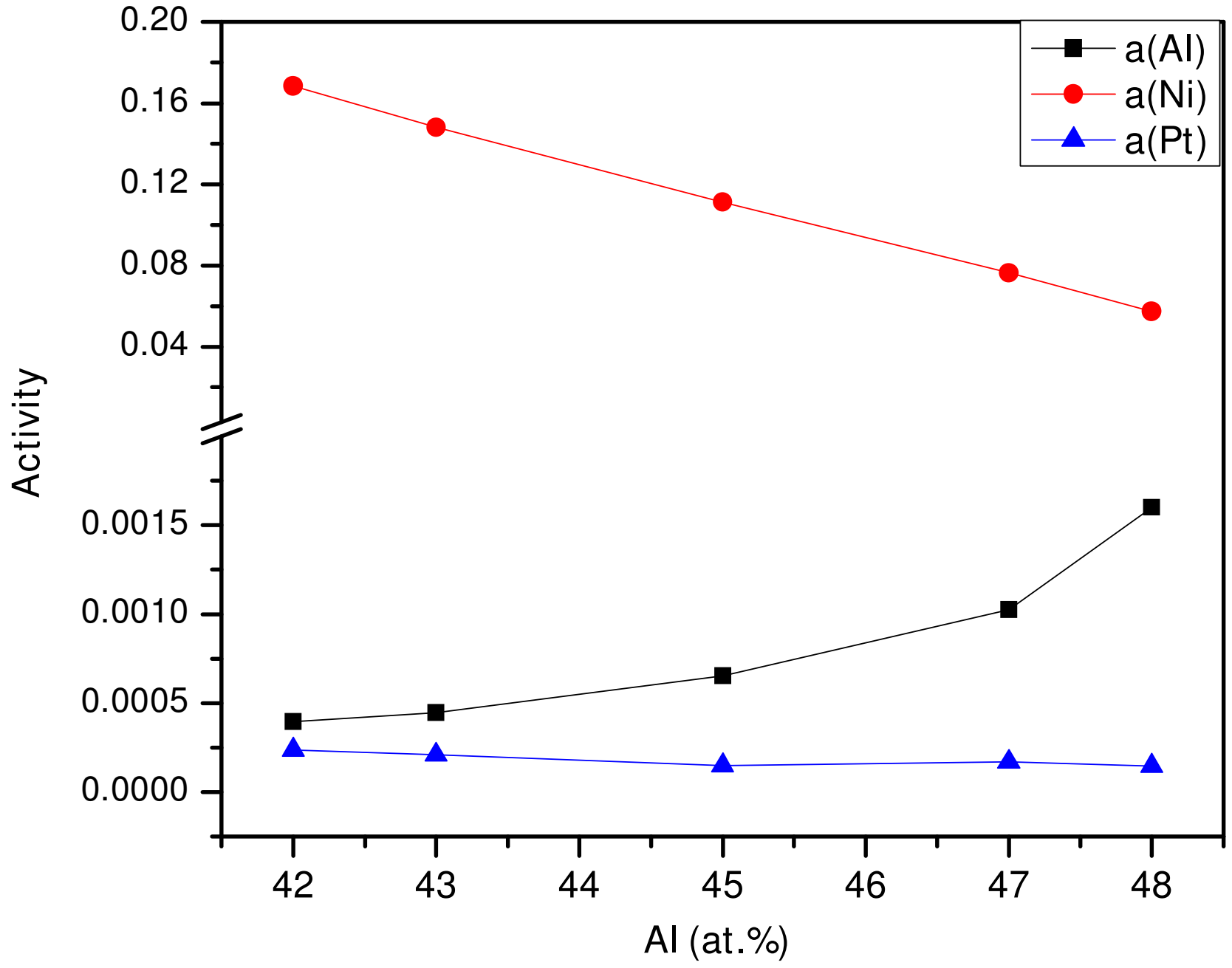

Figure 2 Variation of activities in β-Ni(10P)Al/Ni(10Pt)50Al diffusion couple extracted using CALPHAD.

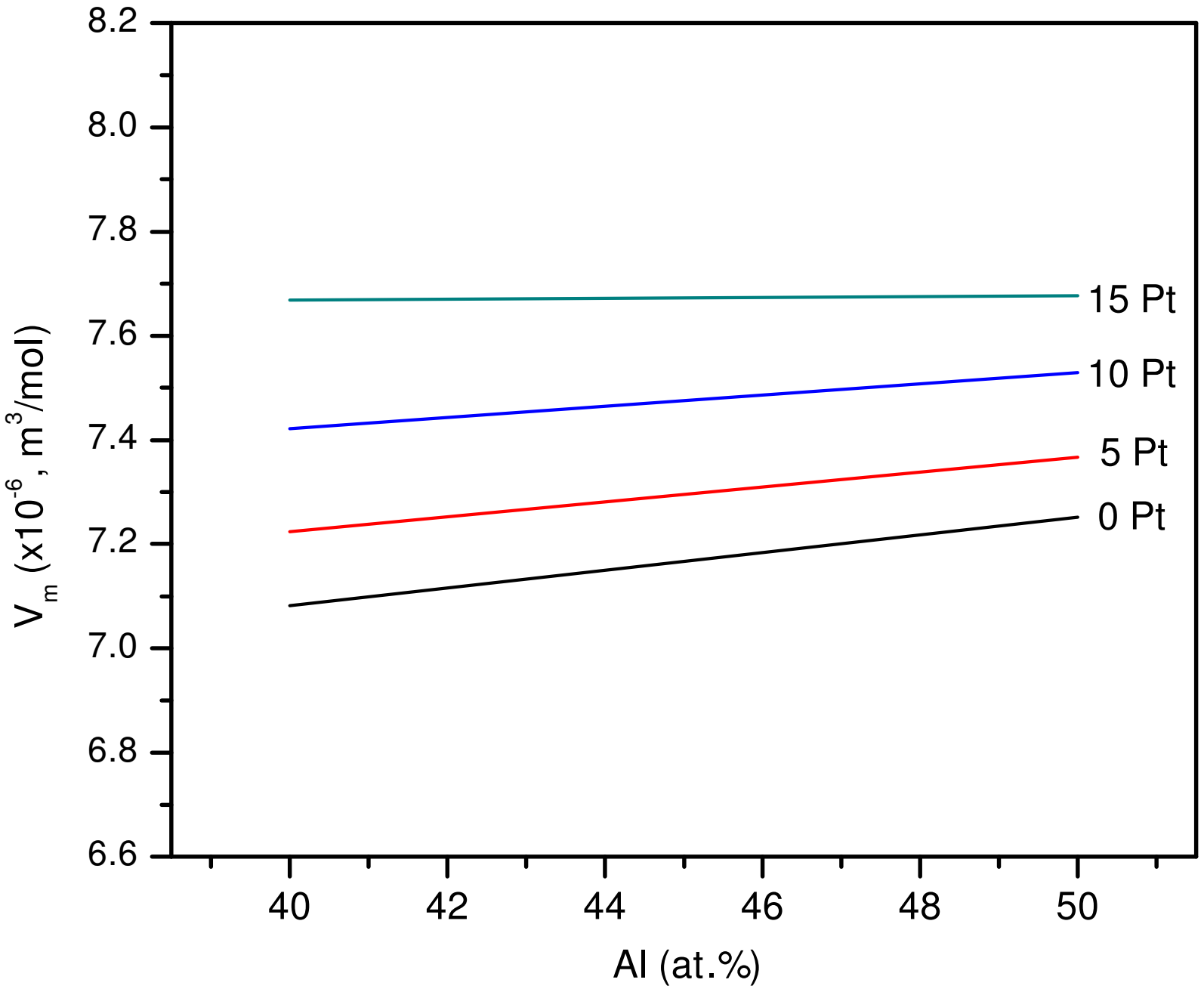

Figure 3: Variation of molar volume with Pt in β-NiPtAl.

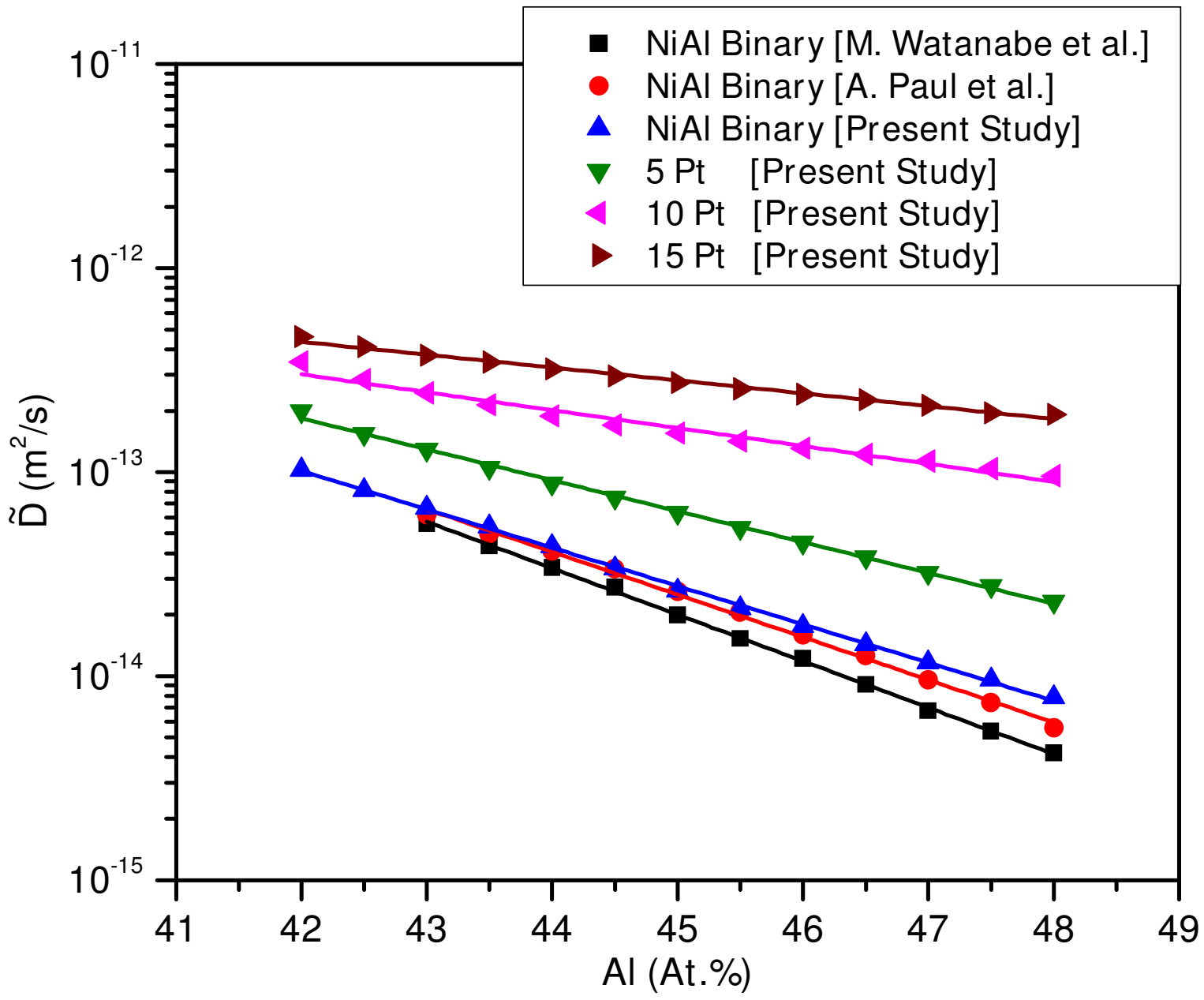

(a)

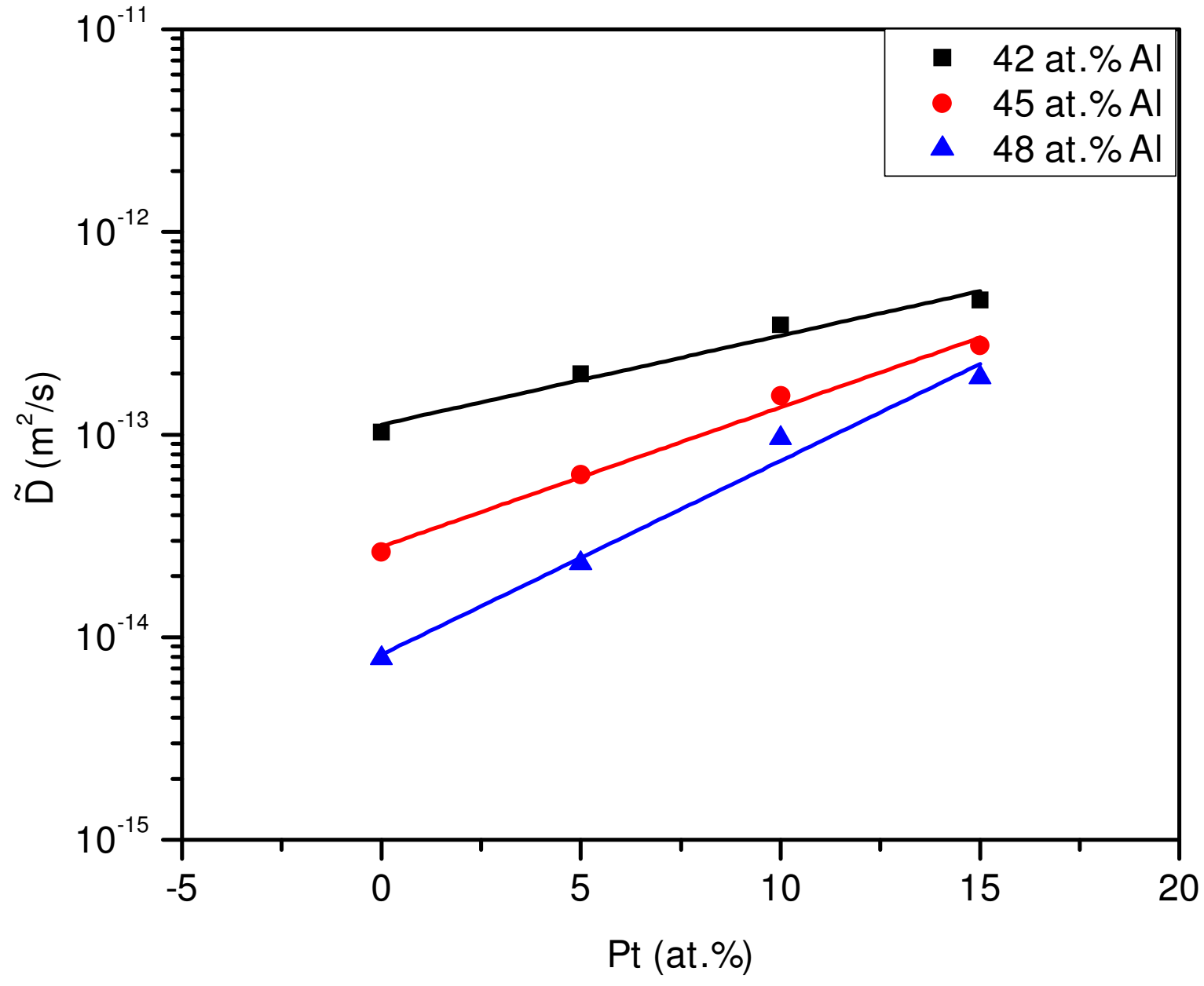

(b)

Figure 4: a) Variation of interdiffusion coefficient with Pt in β-Ni(Pt)Al as a function of Al content. b) Variation of interdiffusion coefficient with Pt in β-Ni(Pt)Al for 42, 45 and 48 at.% Al content.